\documentclass[conference]{IEEEtran}

\usepackage{cite}
\usepackage{graphicx}
\usepackage{subcaption}
\usepackage{booktabs}

\usepackage{algorithm}
\usepackage[noEnd]{algpseudocodex}
\usepackage{amsmath}

\usepackage{color}
\usepackage{soul}
\definecolor{Blue}{cmyk}{0.19, 0.024, 0, 0.012}
\definecolor{Orange}{cmyk}{0, 0.055, 0.19, 0.008}
\definecolor{Red}{cmyk}{0.01, 0.10, 0.03, 0}

\newcommand{\usec}{$\mu$s }

\newcommand*\funcname[1]{\textproc{#1}}

\makeatletter
\def\ps@IEEEtitlepagestyle{%
  \def\@oddfoot{\mycopyrightnotice}%
  \def\@oddhead{}%
  \def\@evenhead{}%
  \def\@evenfoot{}%
}
\def\mycopyrightnotice{%
  \begin{minipage}{\textwidth}
  \centering \scriptsize \copyright~2026 IEEE. Personal use of this
  material is permitted. Permission from IEEE must be obtained for all
  other uses, in any current or future media, including
  reprinting/republishing this material for advertising or promotional
  purposes, creating new collective works, for resale or
  redistribution to servers or lists, or reuse of any copyrighted
  component of this work in other works.
  \end{minipage}
}
\makeatother

\begin{document}

\title{Host-Driven Flowlet Balancing with \\
  Segment Routing over IPv6}

\author{%
  \IEEEauthorblockN{Ryo Nakamura}
  \IEEEauthorblockA{\textit{Information Technology Center}\\
    \textit{The University of Tokyo}\\
    Tokyo, Japan \\
    upa@nc.u-tokyo.ac.jp}
  \and
  \IEEEauthorblockN{Hiroki Kano}
  \IEEEauthorblockA{\textit{InfoTech Div}\\
    \textit{Toyota Motor Corporation}\\
    Tokyo, Japan \\
    hiroki\_kano@mail.toyota.co.jp}
  \and
  \IEEEauthorblockN{Tomoko Okuzawa}
  \IEEEauthorblockA{\textit{InfoTech Div}\\
    \textit{Toyota Motor Corporation}\\
    Tokyo, Japan \\
    okuzawa@mail.toyota.co.jp}
}

\maketitle

\begin{abstract}

  This paper proposes a fully host-driven method for flowlet balancing
  with Segment Routing over IPv6 (SRv6). In modern data center
  networks, load balancing plays a pivotal role in efficiently
  utilizing multiple paths. Flowlet balancing offers finer granularity
  in traffic splitting than ECMP and is therefore expected to achieve
  higher performance. However, deploying flowlet balancing in practice
  is still challenging due to the scalability issue of switches having
  to maintain per-flow state.
  In our approach, hosts detect flowlets in their outgoing traffic and
  steer them onto specific paths using SRv6. The switches behave as
  SRv6 nodes in a stateless manner. Each host distributes its flowlets
  as evenly as possible across paths. As a metric for this load
  balancing, we introduce a simple model that estimates in-flight
  bytes on each path.
  We implemented the proposed method on Linux and evaluated it on a
  testbed with an SRv6-capable router. The results show that, under
  fixed-size flows, the proposed method reduces tail latency by 15\%
  and 33\% compared with random flowlet balancing and ECMP,
  respectively. Furthermore, combining the method with dynamically
  adjusted flowlet timeouts also improves performance under two
  application workloads.

\end{abstract}

\begin{IEEEkeywords}
  Load balancing, Flowlet, Segment Routing, Data Center Networks
\end{IEEEkeywords}

\section{Introduction}

Modern data center networks adopt Clos topologies, such as spine-leaf
topologies, where multiple paths exist between end hosts (servers) to
scale capacity and handle massive traffic. Balancing traffic across
paths as evenly as possible is a significant issue. Traffic skew can
cause congestion on specific paths, leading to increased tail
latency~\cite{plb}.

When balancing traffic across multiple paths, there are three levels
of granularity: per-packet, per-flow, and per-flowlet. Balancing
traffic across paths per-packet~\cite{rps} offers the finest
granularity and provides near-ideal load distribution; however, it
introduces the possibility of packet reordering, which can degrade
transport-layer performance. Per-flow is the most common approach,
thanks to Equal Cost Multi-path (ECMP) available in commodity
switches. Meanwhile, it is well known that ECMP is load-agnostic. ECMP
selects paths based on the 5-tuple hash values of flows and can cause
traffic skew~\cite{wcmp}.

Flowlet balancing~\cite{flowlet} is the third approach, which
distributes a single flow across multiple paths. For a single flow,
when the interval since sending the last packet to a path exceeds the
delay difference between the multiple paths, the next packet can be
sent on another path without packet reordering.  A flowlet is a burst
of packets delimited by inter-packet gaps that exceed such an
interval, called flowlet timeout value (FTV) or $\delta$. Flowlet
balancing distributes traffic across paths on a per-flowlet basis so
that it achieves finer granularity than per-flow balancing while
reducing packet reordering.


Although flowlet balancing appears to be the optimal approach, it has
not been widely deployed due to its stateful nature. Splitting a flow
into flowlets requires tracking the timestamp of the last packet
sent. Maintaining such per-flow state for millions of flows in a
switch is difficult due to hardware limitations. For example, a
commercial data center switch can hold state for a total of 32,768
flows~\cite{qfx-dlb}; with 32 hosts per switch, this allows only
1,024 flows per host. The recent adoption of flowlet balancing in
AI/ML infrastructures depends on their unique traffic pattern: a small
number of extremely large flows occur
intermittently~\cite{alibabahpn}. Several studies have proposed
methods for flowlet balancing without per-flow state on
switches~\cite{presto, clove, burstloader}; however, these methods
involve complex mechanisms for routing flowlets to specific paths and
detecting congestion to achieve effective load balancing.

To address this issue, we propose a practical host-driven method for
flowlet balancing by using Segment Routing over IPv6
(SRv6)~\cite{rfc8986}. SRv6 is a recent routing mechanism that enables
source routing and is increasingly being adopted in commercial
networking products. In our approach, hosts detect flowlets in their
outgoing traffic and distribute them across spine switches on distinct
paths; the spine switches act as stateless SRv6 nodes. The hosts
distribute their flowlets across paths as evenly as possible. As a
metric for this balancing, we introduce a simple model for hosts to
estimate the number of in-flight bytes on a path. These techniques are
self-contained; they run inside each host autonomously. Our approach
requires no complex mechanisms for routing and congestion detection
and thus brings operational simplicity.

We implemented the proposed method on Linux and evaluated it on a
testbed built with a commercial SRv6 router. The results demonstrate
that balancing based on in-flight byte estimation reduces the 99th
percentile flow completion time (FCT) by 15\% and 33\% compared with
random balancing and ECMP, respectively. Moreover, results from two
different workloads show that the method with dynamic FTV adjustment
halves tail FCTs for large flows compared with ECMP.

The contributions of this paper include the following:
\begin{itemize}
\item We propose a method for fully host-driven flowlet balancing by
  using Segment Routing over IPv6.
\item We introduce a model for estimating in-flight bytes from hosts
  as a metric for load distribution.
\item The evaluation on a physical testbed demonstrates that balancing
  based on in-flight byte estimation can shorten flow completion time.

\end{itemize}

\section{Related Work}

Flowlet balancing was originally introduced to split a single TCP flow
across multiple paths in networks~\cite{flowlet}, and has been adopted
for data center networks, where multiple paths are common. There are
two main directions for improving flowlet balancing performance: (1)
balancing algorithms and (2) FTV tuning. The former refers to how to
select paths for each flowlet. CONGA~\cite{conga} and HULA~\cite{hula}
leverage real-time congestion information to select
paths. LetFlow~\cite{letflow} showed that choosing paths randomly can
achieve good performance, with fixed FTVs. On the other hand, recent
studies have addressed adjusting FTVs dynamically on a per-flow
basis~\cite{flex, halflife}. In particular, Halflife~\cite{halflife}
experimentally demonstrated that longer FTVs are suitable for short
flows, whereas shorter FTVs are suitable for long flows.

An emerging use case of flowlet is AI/ML infrastructure. AI/ML
workloads involve Remote Direct Memory Access (RDMA) communication
among many GPUs, with a small number of elephant flows (typically
fewer than several hundred per host, each at over 100~Gbps) that occur
intermittently~\cite{alibabahpn}. This extreme use case offsets the
drawback of flowlet balancing, namely the scalability issue due to
maintaining per-flow state in switches. Merchant switches support
flowlet balancing for this purpose~\cite{dlb}, and recent studies seek
to further improve the performance of flowlet balancing in RDMA
networks~\cite{hf2t, rocelet}.

Host-based flowlet balancing without per-flow state in switches has
also been explored~\cite{presto, burstloader, clove}. Instead of
switches, hosts detect and distribute flowlets over paths. As a
result, these approaches rely on complex mechanisms to route flowlets
to specific paths and detect congestion. Presto~\cite{presto} uses
labels, e.g., MAC addresses or MPLS labels, in a non-standard manner
to encode multiple spanning trees in a
network. BurstLoader~\cite{burstloader} leverages a similar encoding
scheme~\cite{xpath}. CLOVE~\cite{clove} uses an overlay technique but
requires discovering combinations of packet header fields that steer
packets to specific paths. Furthermore, to detect congestion, they
rely on RTT measurements~\cite{burstloader} or ECN~\cite{clove}.

\section{Approach}

To practically eliminate per-flow state from switches for scalability,
we propose using SRv6. Each host in a data center network detects
flowlets in its outgoing traffic and steers them to specific paths
with SRv6. Furthermore, each host distributes its traffic across all
available paths as evenly as possible based on the estimation of
in-flight bytes.

\subsection{Flowlet Balancing with SRv6}

Segment Routing (SR) is a source-based routing mechanism. SR
represents any topological entities, such as nodes, links, and
adjacencies, as \textit{segments}. SR-capable network devices forward
packets in accordance with segment identifiers (SIDs) embedded in the
packets. SRv6 is an instantiation of SR that uses IPv6 addresses as
SIDs and IPv6 encapsulation for packet transport. Currently, it is
supported by commercial routers and is being adopted in data center
networks~\cite{sonic-usid}.

\begin{figure}[t]
  \centering
  \includegraphics[width=1.0\linewidth]{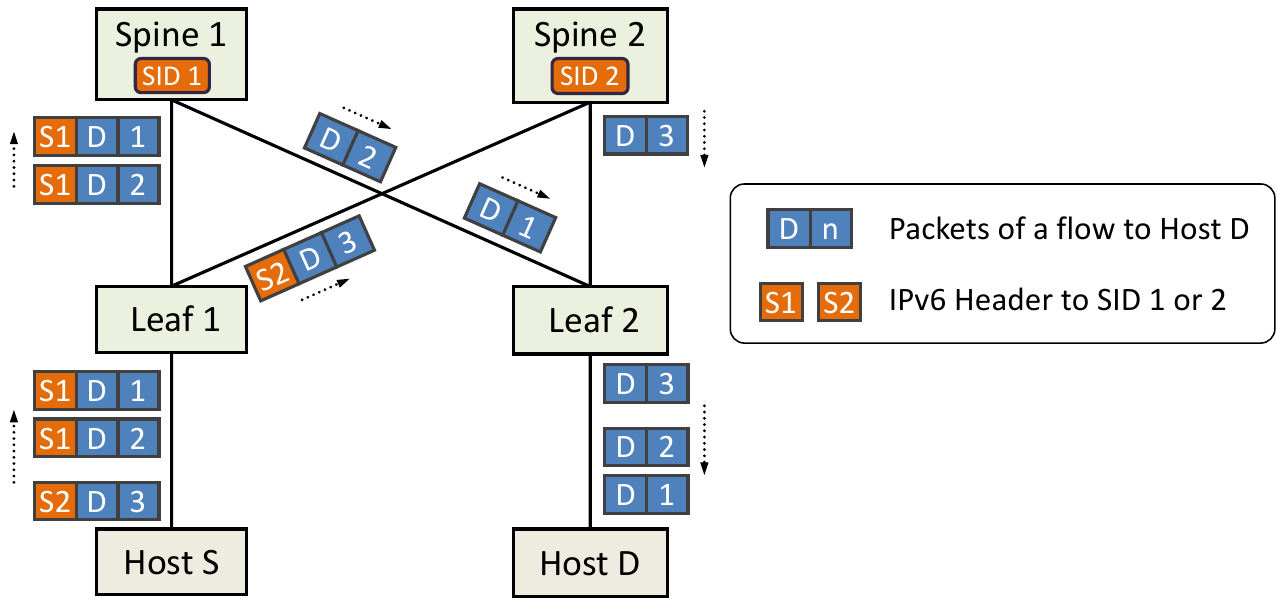}
  \caption{Flowlet balancing with SRv6. The host detects flowlets in
    its outgoing traffic and encapsulates packets with an IPv6 header
    whose destination address is set to the SID of a spine switch.}
  \label{fig:srflb}
\end{figure}

Figure~\ref{fig:srflb} illustrates how host-driven flowlet balancing
with SRv6 works. Each spine switch, which corresponds to a path, has a
unique SRv6 SID. The hosts can steer a packet to a spine switch by
encapsulating it with an IPv6 header whose destination address is the
corresponding SID. When the spine switch receives the packet, the
switch decapsulates it and forwards the inner packet according to its
destination address.\footnote{This behavior is called End with
Ultimate Segment Decapsulation~\cite{rfc8986}.}
Algorithm~\ref{alg:encap} describes the per-packet processing on the
hosts. The hosts maintain per-flow state, namely the timestamp of the
last packet sent for a flow. When a host sends a packet, if the time
since the last packet of the flow exceeds $\delta$ seconds, the host
considers it a new flowlet and selects a new SID. Finally, the packet
is encapsulated in an IPv6 header, with the destination address set to
the SID of a spine switch.

\begin{algorithm}
  \caption{Per-packet processing on hosts}
  \label{alg:encap}
  \begin{algorithmic}[1]
    \small
    \State \textbf{Parameters:}
    \State \hspace{\algorithmicindent} $pkt$: the packet to be sent
    \State \hspace{\algorithmicindent} $now$: the current time
    \State \hspace{\algorithmicindent} $\delta$: the flowlet timeout value
    \State \hspace{\algorithmicindent} $sids$: the SIDs of available spine switches
    \Statex ~

    \Function{EncapsulatePacketForFlowlet}{\textit{pkt}}

    \State $ flow \gets \Call{FindFlow}{\textit{pkt}} $
    \LComment{$flow.last$ is 0 if it is a new flow}

    \If{ $ now - flow.last > \delta$}
    \LComment{A new flowlet, select a SID for it}
    \State $ flow.sid \gets \Call{SelectSID}{sids}$
    \EndIf

    \State $ flow.last \gets now $
    \State $\Call{SRv6Encap}{pkt, flow.sid}$

    \EndFunction
  \end{algorithmic}
\end{algorithm}

In Algorithm~\ref{alg:encap}, there are two variables, $\delta$ and
$sids$, and an unspecified function, \funcname{SelectSID}. The flowlet
timeout value $\delta$ typically ranges from $3\times$RTT to $5
\times$RTT according to prior work~\cite{halflife}. We describe the
values used in our experiment in Section~\ref{sec:exp}. Each SID in
$sids$ corresponds to a spine switch on an available path to the
destination hosts. These SIDs can be provided to the hosts through
configuration, or the hosts can learn them via dynamic routing
protocols, e.g., IS-IS~\cite{rfc9352} and OSPFv3~\cite{rfc9513}.  The
\funcname{SelectSID} function is the heart of load balancing. A simple
algorithm is to randomly assign a SID to every flowlet, as introduced
by LetFlow~\cite{letflow}. The next section describes our approach to
balancing traffic more evenly.

SRv6 encapsulation involves the overhead of a 40-byte outer IPv6
header, and adopting Compressed SID (CSID)~\cite{rfc9800} can
eliminate this overhead. CSID is a 16-bit SID format that can embed
multiple CSIDs within a 128-bit IPv6 address. When using CSID in our
approach, there are two types of CSIDs: one for spine switches and one
for hosts. Applications treat a host CSID as the IPv6 address of the
destination host. The \funcname{SRv6Encap} function in
Algorithm~\ref{alg:encap} inserts a spine-switch CSID into the
destination field of the IPv6 header of outgoing packets. The spine
switch pops its own CSID from the received packets and forwards them
to their original destinations.\footnote{This behavior is called End
with NEXT-CSID~\cite{rfc9800}.}

\subsection{Estimating In-flight Bytes for Load Balancing}

Balancing algorithms select paths for flowlets based on some metric to
avoid traffic skew. Switch-based balancing can employ actual
congestion information, e.g., queue length on egress ports; however, a
host-driven method cannot do so. As a metric, we propose using the
amount of in-flight bytes on each path: the bytes that a host has sent
but have not yet reached the destination hosts. If each host
distributes its flowlets to keep in-flight bytes across paths as
balanced as possible, the overall load across the paths in the network
will be balanced.

\begin{figure}[t]
  \centering
  \includegraphics[width=0.62\linewidth]{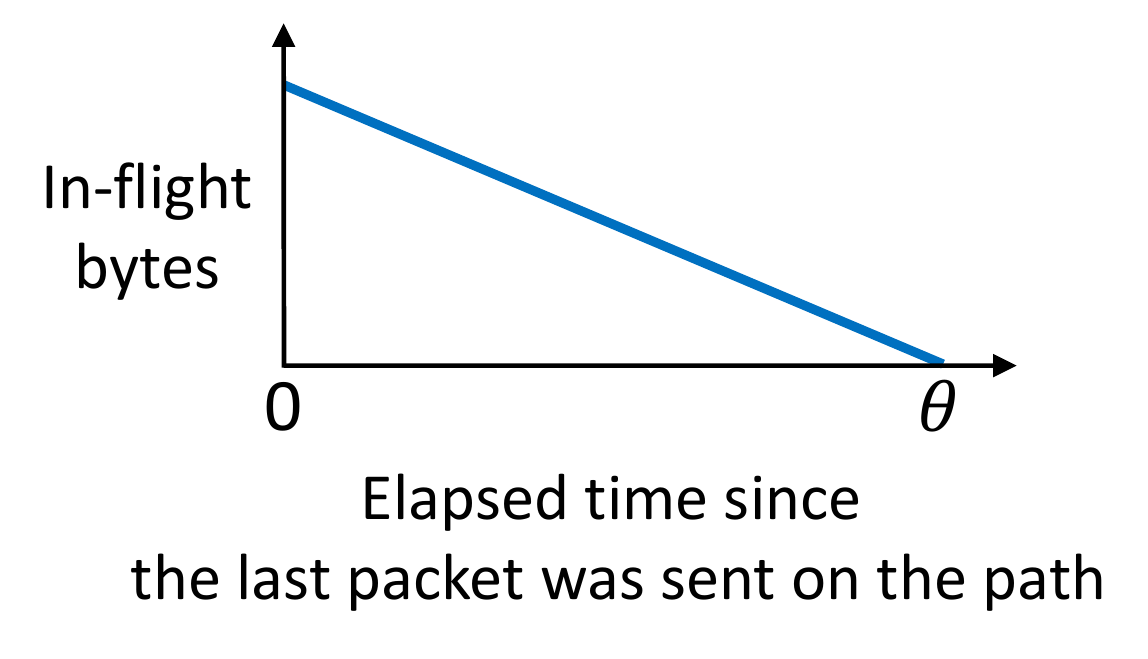}
  \caption{A model for estimating in-flight bytes on a path as a
    function of elapsed time since the last packet was sent.}
  \label{fig:model}
\end{figure}

Accurately measuring in-flight bytes is difficult; instead, we
introduce a simple model to estimate them. Figure~\ref{fig:model}
represents the model. As time elapses, in-flight packets drain from a
path. $\theta$ denotes the time required for all in-flight packets on
the path to drain since the last packet was transmitted.  The
estimated in-flight bytes decrease linearly as time elapses until
$\theta$. The following formula describes the model to calculate
current in-flight bytes, $inflight_{now}$, at the time $T_{now}$,
\begin{equation}
  \small
  inflight_{now} = inflight_{last} \times
  (1 - \frac{min(T_{now} - T_{last}, \theta )}{ \theta }),
\end{equation}
where $inflight_{last}$ is the calculated in-flight bytes at the time
$T_{last}$. The fraction $(T_{now} - T_{last}) / \theta$ represents
how much of the in-flight bytes should have drained during the elapsed
time since $T_{last}$, and it reduces $inflight_{last}$
proportionally. When the elapsed time exceeds $\theta$, the in-flight
bytes are assumed to have drained, and thus, $inflight_{now}$ becomes
zero. The drain timeout value $\theta$ varies depending on the
network. The minimum value must be half of the RTT, i.e., the one-way
delay, but queuing delays increase it. In this study, we set $\theta$
to $10 \times$RTT, based on heuristic exploration.

Algorithm~\ref{alg:p2c} shows an extended version of
Algorithm~\ref{alg:encap} with in-flight byte estimation. The number
of in-flight bytes is maintained per SID representing a path and
updated for every outgoing packet (Lines 7--11 and 34 in
Algorithm~\ref{alg:p2c}). Here, we leverage the power-of-two choices
(P2C) strategy~\cite{p2c}, as in prior work~\cite{halflife}. The
\funcname{SelectSidPowerofTwo} function randomly picks two SIDs in
addition to the current SID and chooses the one with the smallest
estimated in-flight bytes.

\begin{algorithm}
  \caption{Per-packet processing with power-of-two choices
    balancing based on in-flight byte estimation}
  \label{alg:p2c}
  \begin{algorithmic}[1]
    \small
    \State \textbf{Parameters:}
    \State \hspace{\algorithmicindent} $pkt$: the packet to be sent
    \State \hspace{\algorithmicindent} $now$: the current time
    \State \hspace{\algorithmicindent} $\delta$: the flowlet timeout value
    \State \hspace{\algorithmicindent} $\theta$: the drain timeout value
    \State \hspace{\algorithmicindent} $sids$: the SIDs of available spine switches
    \Statex ~

    \Function{EstimateInFlight}{\textit{sid}}
    \State \Return $sid.inflight \times (1 - \dfrac{min(now - sid.last, \theta)}{\theta}) $
    \EndFunction
    \Statex ~

    \Function{UpdateInFlight}{\textit{sid, pktsize}}
    \State $ sid.inflight \gets \Call{EstimateInFlight}{sid} + pktsize $
    \State $ sid.last \gets now $
    \EndFunction
    \Statex ~

    \Function{SelectSidPowerofTwo}{\textit{sids}, $sid_{cur}$}
    \State $ sid_1 \gets \Call{SelectSidRandomly}{sids} $
    \State $ sid_2 \gets \Call{SelectSidRandomly}{sids} $

    \State $ i_1 \gets \Call{EstimateInflight}{sid_1} $
    \State $ i_2 \gets \Call{EstimateInflight}{sid_2} $
    \State $ i \gets \Call{EstimateInflight}{sid_{cur}} $

    \LComment{Select the SID with the smallest in-flight bytes}

    \State $ sid \gets sid_{cur} $
    
    \If{$ i_1 < i $}
    \State $ sid \gets sid_1 $
    \State $ i \gets i_1 $
    \EndIf

    \If{$ i_2 < i $}
    \State $ sid \gets sid_2 $
    \EndIf

    \State \Return $sid$
    \EndFunction
    \Statex ~

    \Function{EncapsulatePacketForFlowlet}{\textit{pkt}}

    \State $ flow \gets \Call{FindFlow}{\textit{pkt}} $
    \LComment{$flow.last$ is 0 if it is a new flow}

    \If{ $ now - flow.last > \delta$}
    \LComment{A new flowlet, select a SID for it}
    \State $ flow.sid \gets \Call{SelectSidPowerofTwo}{sids, flow.sid}$
    \EndIf

    \State $ flow.last \gets now $
    \State $\Call{SRv6Encap}{pkt, flow.sid}$
    \State $ \Call{UpdateInFlight}{flow.sid, pkt.size} $

    \EndFunction
  \end{algorithmic}
\end{algorithm}

The key advantage of our approach is that routing flowlets with SRv6
and estimating in-flight bytes run autonomously inside each
host. There are no centralized controllers or complicated congestion
detection. The spine switches operate as SRv6 nodes without
maintaining per-flow state. We argue that the proposed method is
practical and offers operational simplicity for deploying flowlet
balancing.

\subsection{Implementation}

We implemented the proposed method as a single extend Berkeley Packet
Filter (eBPF) program attached to tc hook in Linux~\cite{tcebpf}. The
program processes every outgoing packet as described in
Algorithm~\ref{alg:p2c}. Per-flow state is maintained in an eBPF map,
an in-memory data structure. Available SIDs are passed from external
sources, e.g., manual configuration or routing daemons, to the program
via an eBPF map. The implementation is capable of both IPv6
encapsulation and CSID insertion, and we adopt the CSID version in the
rest of the paper because it incurs no encapsulation overhead.

In addition to the algorithm described above, we implemented four
balancing methods, listed in Table~\ref{tab:comp}, all running over
SRv6. RPS is random packet spraying that treats every packet as a new
flowlet with an FTV of 0 and randomly selects SIDs. LetFlow uses a
fixed FTV and selects SIDs at random. P2C is the power-of-two choices
algorithm with in-flight byte estimation, as described in
Algorithm~\ref{alg:p2c}. We also implemented the fading module of
Halflife~\cite{halflife}, which dynamically adjusts FTVs on a per-flow
basis. Halflife-RND uses the fading module and randomly selects SIDs,
while Halflife-P2C uses the fading module and P2C to select SIDs.

\begin{table}[h]
  \renewcommand{\arraystretch}{1.3}
  \centering
  \caption{Balancing algorithms we implemented over the SRv6-based flowlet
    routing mechanism.}
  \label{tab:comp}
  \begin{tabular}{c|c|c}
    \toprule
    Method           & FTV      & Balancing Algorithm \\
    \midrule
    RPS            & 0        & Random    \\
    LetFlow        & Fixed    & Random    \\
    P2C            & Fixed    & P2C with in-flight byte estimation \\
    Halflife-RND   & Dynamic  & Random    \\
    Halflife-P2C   & Dynamic  & P2C with in-flight byte estimation \\
    \bottomrule
  \end{tabular}
\end{table}

\section{Testbed Evaluation}
\label{sec:exp}

We conducted an experiment to compare the methods listed in
Table~\ref{tab:comp} and ECMP on a physical testbed. The evaluation
metric is flow completion time (FCT). Better load balancing
distributes traffic across paths more evenly, avoids traffic skew and
packet drops, and thereby alleviates tail latency.

\subsection{Experimental Setup}

\begin{figure}[t]
  \centering
  \begin{minipage}{0.67\linewidth}%
    \centering
    \includegraphics[width=1.0\linewidth]{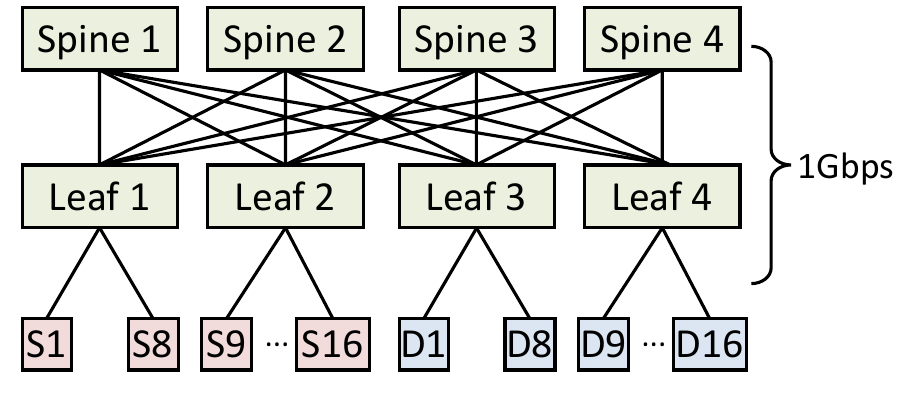}
    \subcaption{Virtual topology}
    \label{fig:tb:v}
  \end{minipage}
  \begin{minipage}{0.3\linewidth}%
    \centering
    \includegraphics[width=1.0\linewidth]{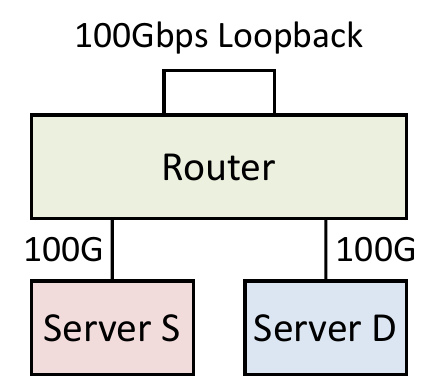}
    \subcaption{Physical setup}
    \label{fig:tb:p}
  \end{minipage}
  \caption{The testbed structure.}
  \label{fig:tb}
\end{figure}

Figure~\ref{fig:tb} depicts our testbed. The experimental topology is
shown in Figure~\ref{fig:tb:v}; there are four leaf switches, and each
is connected to eight hosts and four spine switches with 1~Gbps
links. Namely, it is a 2:1 over-subscription. This topology was
realized on the physical setup shown in Figure~\ref{fig:tb:p}. The
router was MX10003 from Juniper Networks that supports partitioning
interfaces and control planes into independent routing entities,
called logical systems. All spine and leaf switches were deployed as
logical systems. Two servers, each having two Intel Xeon 5318Y CPUs
and an NVIDIA ConnectX-6 100~Gbps NIC, were connected to the router
with a 100~Gbps link each. Each server ran 16 containers that emulated
independent hosts having an SR-IOV VF, and they were pinned to
different CPU cores. All links between the switches and hosts were
multiplexed with VLANs on the 100~Gbps links. The router set the
bandwidth of these VLAN sub-interfaces to 1~Gbps with its QoS feature.
In addition, the containers running FRRouting~\cite{frr} dynamically
exchanged routes and SIDs with all switches via IS-IS.

We used flowperf~\cite{flowperf}, a flow-based benchmarking tool, to
generate flows. In each measurement run, each of the 16 source hosts
(\texttt{S1} to \texttt{S16}) generated flows toward their respective
destination hosts (\texttt{D1} to \texttt{D16}) for 300 seconds. Each
host transmitted four flows concurrently, and the transmission rate
was limited to 500~Mbps using the TX rate limiting of the SR-IOV
VFs. Consequently, a leaf switch received $8\times$500~Mbps of
incoming traffic, which was balanced across four 1~Gbps paths. If
traffic is perfectly balanced, no packet drops occur; however, when
skew arises, congestion builds up in the switch queues, resulting in
increased FCTs and packet drops. Note that we assumed the RTT of the
network was 100 \usec based on measurements; therefore, FTV $\delta$
was set to 500 \usec for LetFlow and P2C, and $\theta$ was set to 1
ms.

\subsection{Results with a Simple Workload}

\begin{figure}[t]
  \centering
  \includegraphics[width=0.9\linewidth]
  {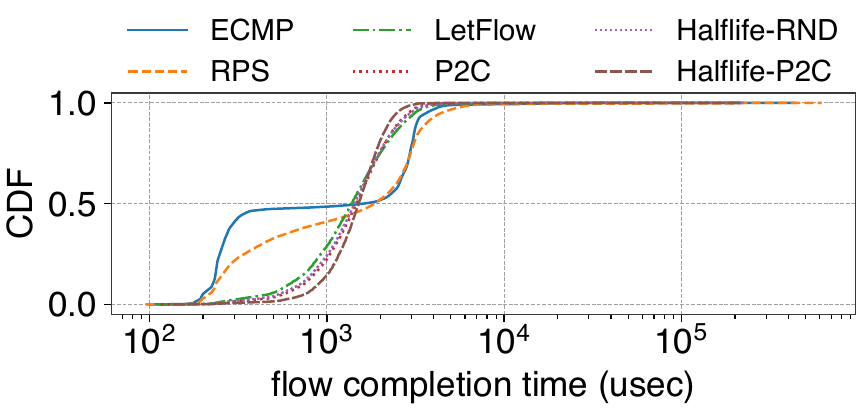}
  \caption{CDF of FCTs with 100KB flows.}
  \label{fig:100k:fct}
\end{figure}

First, we ran the test with 100KB flows to examine a baseline
performance. Figure~\ref{fig:100k:fct} shows the CDF of FCTs for each
method. The result of ECMP shows a step-like pattern. When traffic
becomes skewed on specific paths, those paths are congested while
others are not. As a result, FCTs for flows of the same size vary
depending on which paths they traverse. The ECMP result reflects this
imbalance. In contrast, flowlet balancing distributes traffic more
evenly, resulting in less variable FCTs and more vertically aligned
CDF curves.

\begin{figure}[t]
  \centering
  \includegraphics[width=0.9\linewidth]
  {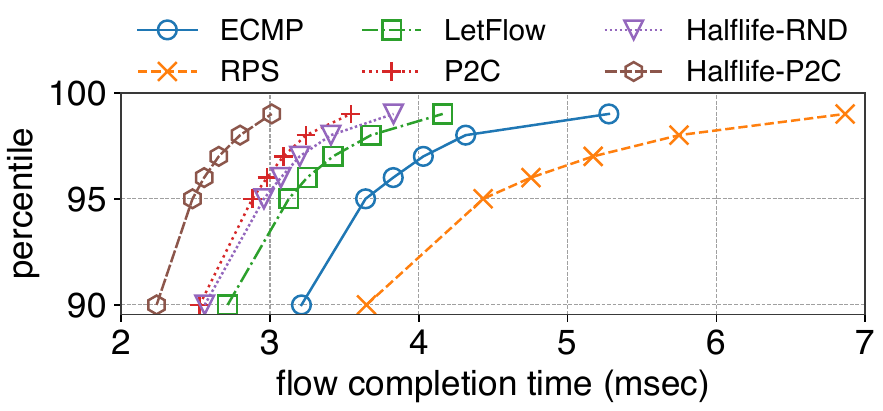}
  \caption{Tail FCTs with 100KB flows.}
  \label{fig:100k:tail}
\end{figure}

Figure~\ref{fig:100k:tail} shows the 90th and 95th--99th percentile
FCTs from Figure~\ref{fig:100k:fct}. As shown, SRv6-based flowlet
balancing reduces the tail FCTs compared with ECMP. For the 99th
percentile FCT, P2C achieves 3.54 ms, a 15\% reduction compared with
LetFlow (4.16 ms) and a 33\% reduction from ECMP (5.28 ms). Combining
P2C with the dynamic FTV of Halflife (Halflife-P2C) further improves
performance, reducing it to 3.01 ms, 28\% and 43\% reductions compared
with LetFlow and ECMP, respectively. These results indicate that the
power-of-two choices algorithm with in-flight byte estimation achieves
better balancing than the random path selection. In addition, dynamic
FTV adjustment improves the performance compared with the fixed FTV,
and RPS results in longer tail latency.

\subsection{Results with Realistic Workloads}

Next, we evaluated the methods with two workloads of different
flow-size distributions: Meta Hadoop~\cite{fb-hadoop} and Web
Search~\cite{dctcp}. The former consists mainly of flows from 10B to
1MB, and the latter spans from 1KB to 30MB.

\begin{figure}[t]
  \centering
  \begin{minipage}{1.0\linewidth}%
    \centering
    \includegraphics[width=1.0\linewidth]
    {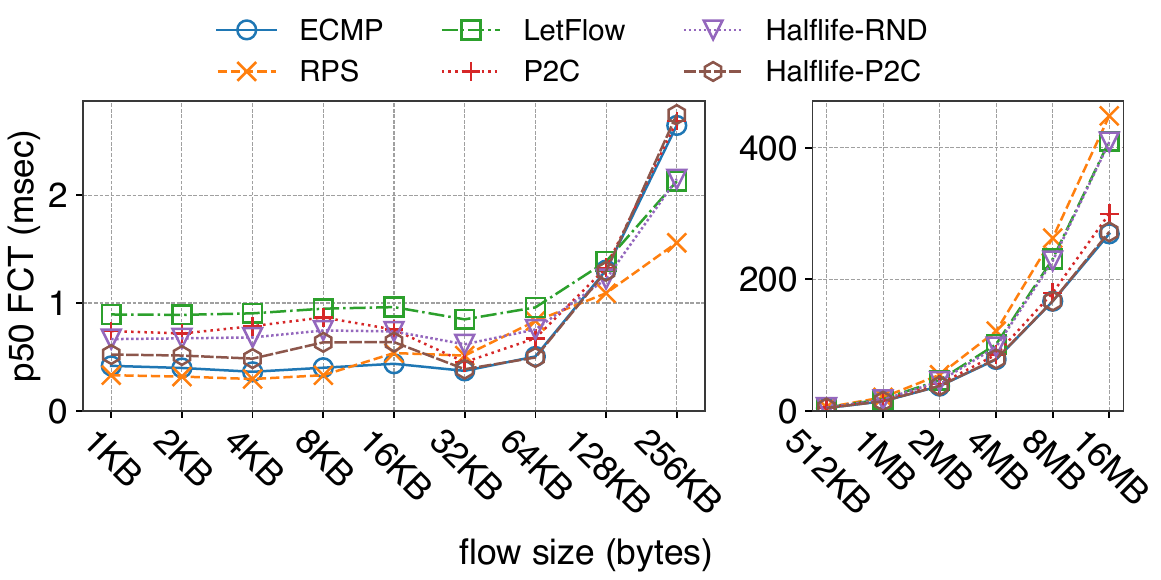}
    \caption{Median FCTs for the Meta Hadoop workload.}
    \label{fig:fb:50}
  \end{minipage}
  \begin{minipage}{1.0\linewidth}%
    \centering
    \includegraphics[width=1.0\linewidth]
    {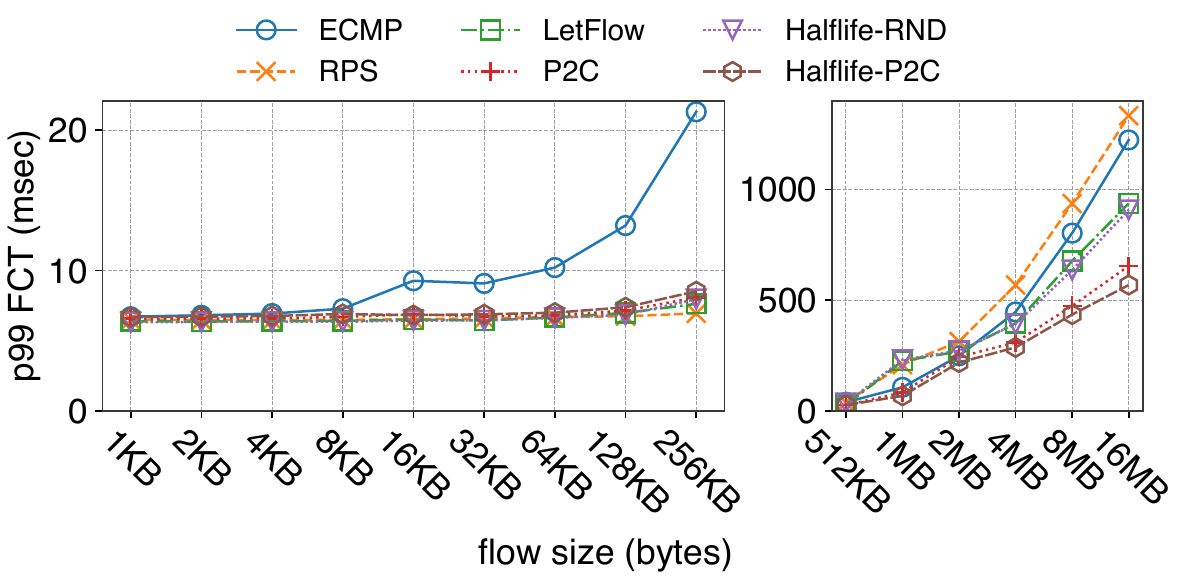}
   \caption{99th-percentile FCTs for the Meta Hadoop workload.}
    \label{fig:fb:99}
  \end{minipage}
\end{figure}

Figure~\ref{fig:fb:50} shows the median FCT with the Meta Hadoop
workload. The x-axis represents flow size ranges: 1KB indicates flows
of $($0, 1KB$\rbrack$, 2KB indicates flows of $($1KB, 2KB$\rbrack$,
4KB indicates $($2KB, 4KB$\rbrack$, and so on. For the median FCT,
ECMP performs the best. This result is also due to traffic skew; flows
traversing uncongested paths complete quickly. Flowlet balancing, in
contrast, evenly congests all paths. Nevertheless, compared to the
other methods, Halflife-P2C achieved FCTs close to ECMP except for the
256KB range.

Figure~\ref{fig:fb:99} shows the 99th percentile FCT under the Meta
Hadoop workload. The tail FCTs are an order of magnitude larger than
the median FCT, and thus reducing such tail latency has a significant
impact on upper-layer performance.  When the flow size exceeds 4KB,
the tail FCT of ECMP begins to increase, whereas flowlet balancing
effectively suppresses this growth. At a flow size of 256KB,
Halflife-P2C, P2C, and LetFlow reduce the FCT to 8.5 ms, 8.1 ms, and
7.6 ms, respectively, from 21.3 ms with ECMP, achieving approximately
60\% FCT reductions. Although LetFlow slightly outperforms the P2C
methods at 256KB, the P2C methods achieve better performance as the
flow size increases. For the largest flows (8--16MB), the FCT of ECMP
is 1223 ms, while LetFlow reduces it to 936 ms (a 23\% reduction), P2C
reduces it to 654 ms (44\%), and Halflife-P2C reduces it to 568 ms
(54\%). These reductions demonstrate the effectiveness of power-of-two
choices with in-flight byte estimation and the benefit of Halflife's
dynamic FTV adjustment.

\begin{figure}[t]
  \centering
  \includegraphics[width=1.0\linewidth]
  {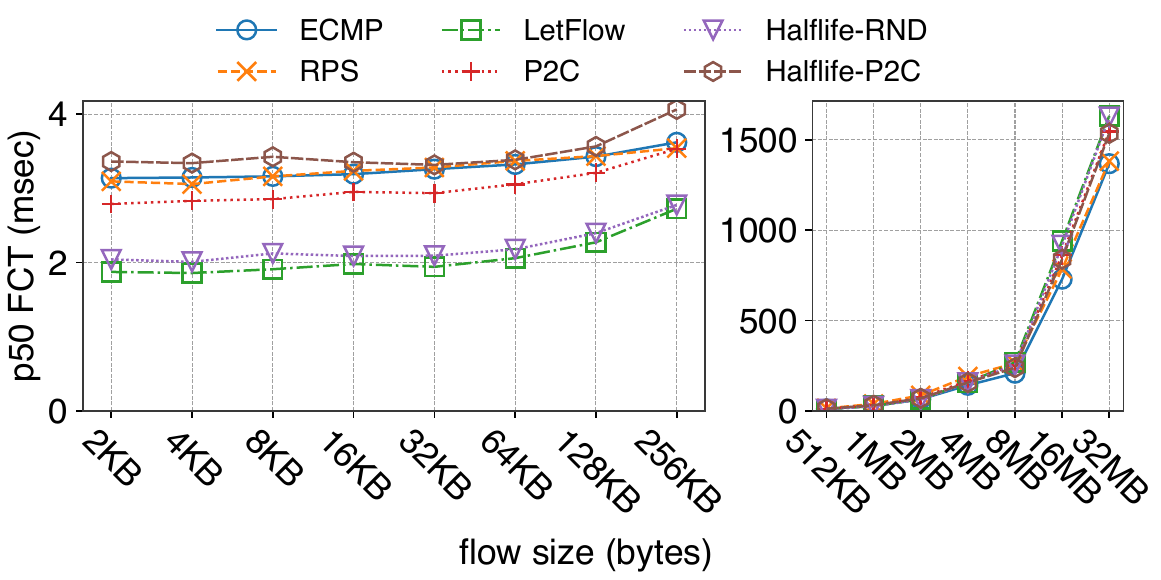}
  \caption{Median FCTs for the Web Search workload.}
  \label{fig:dctcp:50}
\end{figure}

\begin{figure}[t]
  \centering
  \includegraphics[width=1.0\linewidth]
  {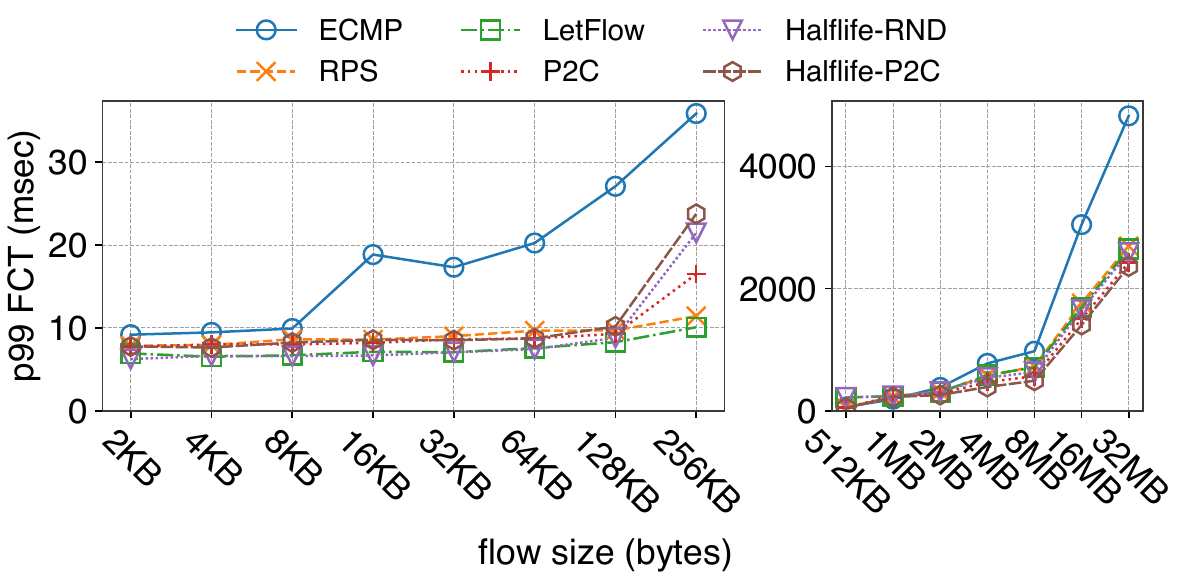}
  \caption{99th-percentile FCTs for the Web Search workload.}
  \label{fig:dctcp:99}
\end{figure}

Figure~\ref{fig:dctcp:50} and \ref{fig:dctcp:99} show the median and
99th percentile FCTs under the Web Search workload. In this workload,
where overall flow sizes are relatively large, LetFlow achieves the
best performance for small flows, up to roughly 1MB. However, as the
flow size increases, Halflife-P2C outperforms the others. For the 99th
percentile FCT of large flows (16--32MB), ECMP yields 4.8 seconds,
while LetFlow reduces it to 2.6 seconds and Halflife-P2C further
reduces it to 2.4 seconds; it approximately halves the tail FCT
compared with ECMP.

\section{Conclusion}

In this paper, we propose a fully host-driven method for flowlet
balancing with SRv6. The hosts detect flowlets in their outgoing
traffic and steer them onto specific paths using SRv6. The hosts
distribute flowlets based on the estimated in-flight bytes on each path as
the metric. These mechanisms are self-contained and run inside each
host autonomously, thereby eliminating per-flow state from switches.
The evaluation on a physical testbed demonstrates that the proposed
method outperforms the others based on random path selection.  In
particular, P2C with dynamic FTV adjustment reduces the 99th
percentile FCT by 28\% and 43\% compared with LetFlow and ECMP,
respectively, under fixed-size flows. The results from two different
workloads show that the method approximately halves the tail FCTs
compared with ECMP for large flows.
As future work, we will analyze the model for in-flight byte
estimation, investigate suitable parameters such as drain timeout
values, and conduct extensive evaluation through simulation.

\bibliographystyle{IEEEtran}
\bibliography{srflb}

\end{document}